\documentclass[sigconf,authorversion,nonacm]{acmart}

\usepackage{booktabs}

\usepackage[utf8]{inputenc}
\usepackage{enumitem}
\usepackage{xcolor}
\usepackage{amsmath}
\usepackage{subfigure}
\usepackage{cleveref}
\usepackage{listings}
\usepackage{fancyvrb}
\usepackage{xurl}
\definecolor{darkgray}{rgb}{0.33, 0.33, 0.33}

\lstnewenvironment{Python}[1][]
  {\lstset{language=Python,
    basicstyle      = \scriptsize\ttfamily,
    keywordstyle    = \color{blue},
    keywordstyle    = [2] \color{teal}, 
    stringstyle     = \color{magenta},
    literate={ü}{{\"u}}1 {ö}{{\"o}}1 {É}{{\'E}}1 {œ}{{\oe}}1,
    deletekeywords=[2]{compile},    
    commentstyle    = \color{darkgray}\ttfamily,
           morekeywords=as,
           morekeywords=with,
           #1}%
  }
  {}

\usepackage{balance}

\newcommand{\header}[1]{\vspace{1mm}\noindent\textbf{#1}.}
\newcommand{\headerl}[1]{\vspace{1mm}\noindent\textit{#1}.}

\newcommand{\github}{\url{https://github.com/stefan-grafberger/shadow-pipeline-experiments}}

\AtBeginDocument{%
  \providecommand\BibTeX{{%
    \normalfont B\kern-0.5em{\scshape i\kern-0.25em b}\kern-0.8em\TeX}}}

\begin{document}

\title{Towards Interactively Improving ML Data Preparation Code via~``Shadow~Pipelines''}

\author{Stefan Grafberger}
\email{s.grafberger@uva.nl}
\orcid{0000-0002-9884-9517}
\affiliation{%
  \institution{AIRLab, University of Amsterdam}
  \country{The Netherlands}
}

\author{Paul Groth}
\email{p.groth@uva.nl}
\orcid{0000-0003-0183-6910}
\affiliation{%
  \institution{University of Amsterdam}
  \country{The Netherlands}  
}

\author{Sebastian Schelter}
\email{schelter@tu-berlin.de}
\orcid{0000-0003-4722-5840}
\affiliation{%
  \institution{BIFOLD \& TU Berlin}
  \country{Germany}  
}

\renewcommand{\shortauthors}{Grafberger, et al.}

\begin{abstract}
Data scientists develop ML pipelines in an iterative manner: they repeatedly screen a pipeline for potential issues, debug it, and then revise and improve its code according to their findings. However, this manual process is tedious and error-prone. Therefore, we propose to support data scientists during this development cycle with automatically derived \textit{interactive suggestions for pipeline improvements}. We discuss our vision to generate these suggestions with so-called \textit{shadow pipelines}, hidden variants of the original pipeline that modify it to auto-detect potential issues, try out modifications for improvements, and suggest and explain these modifications to the user. We envision to apply incremental view maintenance-based optimisations to ensure low-latency computation and maintenance of the shadow pipelines. We conduct preliminary experiments to showcase the feasibility of our envisioned approach and the potential benefits of our proposed optimisations.
\end{abstract}

\maketitle

\section{Introduction}

Software systems that learn from user data with machine learning (ML) have become ubiquitous over the last years and participate in many critical decision-making processes. Unfortunately, real-world experience shows that the pipelines for data preparation, feature encoding, and model training in ML systems are often brittle, especially with respect to issues in the data they process~\cite{breck2019data,polyzotis2018data,stoyanovich2022,schelter2018challenges}. As a consequence, several data-centric techniques are being developed to detect, quantify, and improve ML applications with respect to reliability, fairness, and prediction quality~\cite{jia2019efficient,schelter2021jenga,li2019cleanml,schelter2018deequ,chung2019slice,tramer2017fairtest,bird2020fairlearn}. 
Applying these techniques to ML pipelines currently requires a high level of expertise, as existing approaches such as mlinspect~\cite{grafberger2022mlinspect,grafberger2021demo}, DataScope~\cite{karlavs2023canonpipe}, mlwhatif~\cite{grafberger2023mlwhatif,grafberger2023demo}, Rain~\cite{wu2020complaint}, Gopher~\cite{pradhan2022interpretable} or ArgusEyes~\cite{schelter2022screening,schelter2023demo} assume that data scientists know in advance what kind of errors they are looking for. 

\header{The need for interactively improving ML pipelines} In reality, data scientists typically do not know in advance what pipeline issues to look for, and often ``discover serious issues only after deploying their systems in the real world''~\cite{holstein2019improving}. At development time, data scientists currently have to iteratively screen their pipeline for potential issues, debug these issues, and then revise and improve the pipelines according to their findings. This process is tedious, as it requires repeated manual code re-organisation and re-execution in an environment like a Jupyter notebook.

We argue that ML pipeline development should be accompanied by \textit{interactive suggestions} to improve the pipeline code, similar to code inspections in modern IDEs like IntelliJ~\cite{codeInspectionsIntelliJ} or text corrections in writing assistants like Grammarly~\cite{grammarlyDemo}. For that, we can re-use some techniques from previous work, but are still faced with a \\\\ set of challenges:
$(i)$~We need \textit{low-latency auto-detection} of pipeline improvement opportunities, to seamlessly integrate into the development workflow; $(ii)$ we should identify \textit{pipeline problems spanning several operators}, instead of being artificially limited to screening individual operators one-at-a-time as in e.g., mlinspect~\cite{grafberger2022mlinspect,grafberger2021demo}; $(iii)$ users should receive~\textit{provenance-enabled explanations}~\cite{chothia2016explaining} for detected problems and suggested improvements.

\begin{figure}[t!]
  \centering
  \includegraphics[width=\columnwidth]{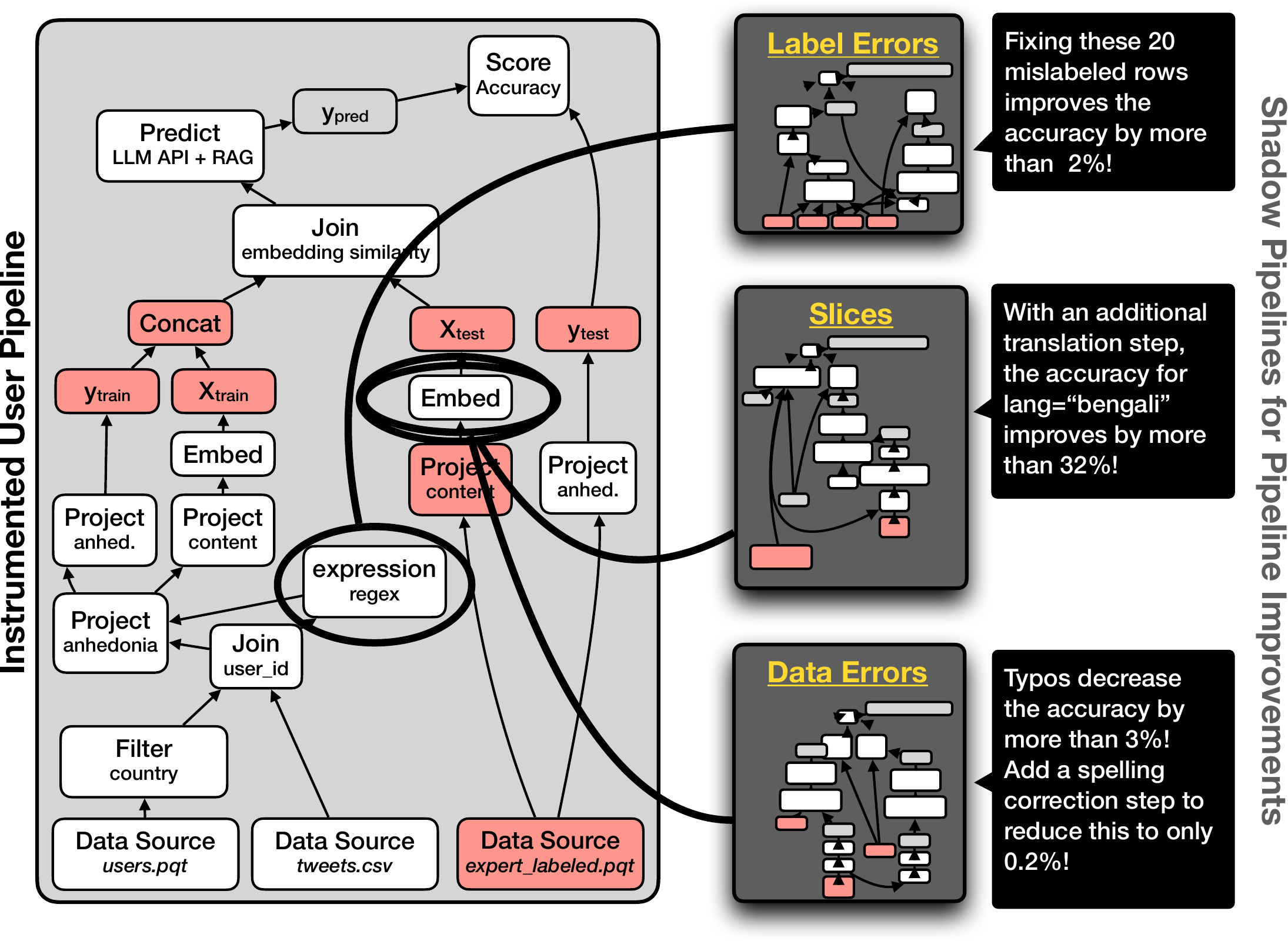}
  \caption{Our vision -- several automatically maintained "shadow pipelines" give actionable suggestions on how to improve a user's ML pipeline code at development time.}
  \label{fig:overview}
\end{figure}

\header{Towards low-latency suggestions for pipeline improvement} We envision a system that instruments a data scientist's ML pipeline code and creates and maintains so-called ``\textit{shadow pipelines}'' with low-latency to generate suggestions for improvements, as illustrated in~\Cref{fig:overview}. Such a shadow pipeline is a hidden variant of the original pipeline, which modifies it to auto-detect potential issues and tries out different pipeline modifications for improvement opportunities. Subsequently, each shadow pipeline provides the user with code suggestions to improve the pipeline, accompanied by a provenance-based explanation and a quantification of the expected impact on the pipeline outputs. From a technical perspective, the main challenge is to conduct the required computations with low latency by reusing and updating intermediates via incremental view maintenance. We introduce the general problem using a running example (\Cref{sec:problem}), present our envisioned approach based on shadow pipelines (\Cref{sec:approach}), and finally conduct preliminary experiments for the feasibility of our approach~(\Cref{sec:evaluation}).
\newpage
In summary, this paper provides the following contributions:
\begin{itemize}[noitemsep,leftmargin=*]
    \item We introduce the problem of interactively generating suggestions during ML pipeline development~(\Cref{sec:problem}).
    \item We propose \textit{shadow pipelines} as an effective approach to computing such suggestions and discuss ideas for efficiently maintaining them based on incremental view maintenance~(\Cref{sec:approach}). 
    \item We conduct preliminary experiments with several shadow pipelines to show the feasibility of computing low-latency suggestions, where we find that our proposed optimisations potentially reduce the runtime by up to two orders of magnitude~(\Cref{sec:evaluation}). 
    \item We provide the source code for our experiments at \textcolor{blue}{\github{}}
\end{itemize}
\section{Problem Statement}
\label{sec:problem}

\header{Running example} Imagine a data scientist who is developing an ML pipeline for a social media platform to detect user posts that show signs of the anhedonia symptom for depression.
\begin{Python}[frame=none,captionpos=b,numbers=left,xleftmargin=.25in,caption={Example pipeline for an NLP use case.},label={lst:example}]
def load_data(countries_to_include):
  users = pd.read_csv('users.csv')
  posts = pd.read_parquet('posts.parquet')  
  users = users[users.country.isin(include_countries)]
  return users.merge(posts, on='user_id')

# Weak labeling with expert rules
def signs_of_anhedonia(post):
  return (post.contains('(0|no|zero) (motivation)')
      | post.contains('lost (interest|motivation)')) 
    & ~(posts.contains('recover from (0|no|zero) interest'))

# Vectorstore retriever for RAG 
def setup_retrieval_corpus(train_data, weak_labels):
  vector_store = Chroma.from_texts(texts=train_data['posts'], 
    metadatas=weak_labels, embedding=...))
  return vectorstore.as_retriever()

# LLM + RAG pipeline using Langchain
def setup_llm_chain():
  return ({"context": retriever | format_docs, 
    "question": RunnablePassthrough()}
    | ChatPromptTemplate.from_template(...)
    | MyCustomLLM()| JsonOutputParser() | format_response)

countries_to_include = ['US', 'CAN']
train_data = load_train_data(countries_to_include)
weak_labels = signs_of_anhedonia(train_data['posts'])
retriever = setup_retrieval_corpus(train_data, weak_labels)
test_data = pd.read_parquet('expert_labeled_data.parquet')
rag_chain = setup_llm_chain(retriever)

y_predicted = rag_chain.batch(test['posts'])
y_test_binarized = label_binarize(test['signs_of_anhedonia'])
accuracy = accuracy_score(y_predicted, y_test_binarized)
\end{Python}

Their pipeline (Listing~\ref{lst:example}) loads and integrates data (Lines~1-5) and performs weak labeling of the training data, based on regular expressions designed by experts (Lines~7-11). Next, the pipeline computes embeddings of the weakly labeled samples, inserts them into a vector store (Lines~13-17), and leverages these embeddings to classify samples using an LLM with retrieval augmentation (Lines~19-24). In lines~26-35, the final pipeline is then evaluated on a test set (which was manually labeled by experts).  Note that this example is inspired by an existing pipeline from NLP research~\cite{nguyen2022depression}. 
\newpage
\header{Iterative improvements for the pipeline} Next, imagine that the data scientist finds that this initial pipeline version provides unsatisfactory accuracy and starts to manually debug and revise the pipeline to figure out how to improve it. They might, for example, write debugging code to look for mislabeled training samples via Shapley Values~\cite{karlavs2023canonpipe} and detect that some wrongly labeled samples are fed to the LLM from the retrieval corpus. 
After mapping the embedded samples back to the original posts, the data scientist would pinpoint the problem to the regular expressions used for weak labeling. They would inspect the problematic posts and extend the regular expressions to cover the cases that were not handled well. Next, the data scientist would re-run the updated pipeline and observe that the accuracy improved a little but that there are still instances of likely mislabeled samples. They again isolate the samples to which the label error detection points and realise that some users' posts contain too many typos to be reliably matched. As a result, the data scientist would once again rewrite the pipeline to integrate a spelling correction step. 
Once no more likely mislabeled samples are detected, the data scientist would try other evaluation techniques, e.g., to find small slices in the test data where the pipeline does not perform well~\cite{chung2019slice,sagadeeva2021sliceline} and start another cycle of iterative debugging and rewriting.

\header{Towards automated, low-latency suggestions for improvements} As already mentioned in the introduction, data scientists spend a lot of time in this iterative process of ``guessing'' what a potential problem might be, rewriting the pipeline to determine if that is indeed the case and then testing and integrating improvements. Furthermore, this tedious process requires a high level of expertise, as the data scientists need to identify the actual improvement opportunities and concrete steps to fix the detected problems.

We argue that we should automate the underlying steps in the outlined iterative improvement cycle as follows. First, common techniques for identifying problematic portions of the data or the data preparation operations (like label error detection or the detection of problematic data slices) should be \textit{automatically run on the pipeline in the background}. Once potential issues are detected, data scientists should be presented with \textit{provenance-based explanations} for the locations in the data or pipeline code that cause the issues. Furthermore, the data scientist should also be presented with a list of \textit{suggestions on how to improve their pipeline}, which ideally already \textit{quantify the expected impact} on a metric of interest such as accuracy or fairness.

Existing systems such as mlwhatif~\cite{grafberger2023demo} or ArgusEyes~\cite{schelter2022screening} fall short to  address this problem, as they assume that the data scientist already knows exactly what issues to look for. They are designed as one-off approaches with no consideration of the iterative development cycle, and in general incur rather high execution times.

\section{Envisioned Approach}
\label{sec:approach}

In the following, we describe our ideas for automatically maintained shadow pipelines to assist the user with ML pipeline development.

\header{Shadow pipelines} Figure~\ref{fig:overview} gives a high-level overview of our envisioned approach, in reference to the running example from \Cref{sec:problem}. On the left, we see an automatically extracted ``logical \newpage plan'' of a user's ML pipeline, consisting of both relational and ML-specific operators. We showed in previous work~\cite{grafberger2022mlinspect,grafberger2023mlwhatif} how to extract such plans by instrumenting Python code. For the user pipeline, we generate so-called shadow pipelines (shown on the right). A shadow pipeline is a hidden variant of the original pipeline, which modifies it to auto-detect potential issues and tries out different pipeline modifications to detect improvement opportunities. Each shadow pipeline consists of several general steps. The first step is issue detection, for which the shadow pipeline introduces operators that take intermediates from the original pipeline as input to screen for potential problems. For detected issues, a shadow pipeline continues with root cause determination, aiming to localise specific pipeline operators and/or input tuples responsible for the identified issue. Finally, the shadow pipeline integrates and evaluates potential fixes, generates the corresponding provenance-based change explanations, and provides a quantification of the expected impact on the pipeline outputs. To prevent information overload, shadow pipelines will have to use various techniques to carefully select tuples for user explanations, e.g., via stratified sampling, slice-finding~\cite{chung2019slice,sagadeeva2021sliceline}, based on Shapley values~\cite{karlavs2023canonpipe,jia2019efficient}, and by focusing on test samples with label changes.

\header{Low-latency computation of shadow pipelines} A key technical challenge is ensuring low-latency shadow pipeline computations for an interactive user experience. For that, we envision to reuse intermediates from the original pipeline, sometimes even on a tuple-level, via incremental view maintenance (IVM) techniques~\cite{mcsherry2013differential}. Every shadow pipeline must only compute the minimum required difference compared to the original user pipeline and should ideally avoid costly operations like model re-training. In cases where such re-training is unavoidable, we plan to explore cheap proxy models~\cite{jia2019efficient} to estimate the impact of a change on expensive models like neural networks. Moreover, certain issue detection techniques and potential fixes can be evaluated on selected subsets of the data only, when faced with costly operations like LLM inference APIs or embedding computations for text or image data. Some shadow pipelines may themselves involve expensive operators like slice finding, Shapley value computation, spelling correction, or text translation. Efficient implementations of these operators will be crucial, and we may even need to work with samples only in extreme cases. Furthermore, the parallel execution of different shadow pipelines and the prioritisation of different analyses will be crucial for performance.

\header{Maintenance of shadow pipelines} During the iterative development cycle of data scientists, they continuously rewrite and re-run their pipelines. In light of such rewrites, we again plan to update the original pipeline and its shadow pipelines with IVM techniques, based on detected changes in operators or input tuples. If the user followed a concrete code suggestion from our shadow pipeline, we may even be able to update the original pipeline with results previously computed in the shadow pipeline. However, in general, updating ML pipelines presents some tough challenges, as IVM approaches~\cite{mcsherry2013differential} typically assume a fixed query with changing input data, while in our scenario, both data and operators can change between iterations. Therefore, we plan to apply a best-effort approach, where we only apply IVM if certain simplifying assumptions are met, e.g. that only a single operator changes.

\headerl{Application to the running example} Consider the exemplary ``label errors'' shadow pipeline shown in Figure~\ref{fig:overview}, which uses Shapley values~\cite{karlavs2023canonpipe,jia2019efficient} to identify likely label errors in the user pipeline's weakly labeled samples. Next, the shadow pipeline simulates fixes in the regexes by flipping the labels of samples with negative Shapley values in the train data, and re-evaluates the pipeline to observe the impact on its accuracy. In the initial shadow pipeline run, we can directly apply the Shapley value algorithm~\cite{karlavs2023canonpipe} on intermediates captured from the original pipeline. Next, we can update the labels of these problematic samples directly in the vector store metadata, and thereby reuse the previously computed train embeddings, which do not change. Finally, the shadow pipeline only needs to rerun the expensive inference for test set predictions that relied on train samples with flipped labels. Similar optimisations can quickly update the shadow pipeline with low latency in case of user changes, and an incremental version of the Shapley value algorithm could even decrease the latency further.
%

\section{Preliminary Experiments}
\label{sec:evaluation}

We present preliminary experiments on computing and maintaining shadow pipelines with hardcoded optimisations. Our goal is to validate the feasibility of continuously managing these pipelines in the background as a user works on their ML pipeline. 

\headerl{User pipelines} We experiment with two versions of the NLP pipeline from our running example in \Cref{sec:problem}. The first version, called \texttt{rag}, uses retrieval augmentation and a trained LLM as detailed in Listing~\ref{lst:example}. The second version, called \texttt{train}, does not use an LLM but trains a simple neural network classifier on the embeddings.

\headerl{Shadow pipelines} We evaluate three shadow pipelines inspired by the issues outlined in~\Cref{sec:problem}: \texttt{slices} uses SliceLine~\cite{sagadeeva2021sliceline} to detect subsets of the test set where the user pipeline does not work well. These samples correspond to posts in non-English languages; therefore, \texttt{slices} integrates an additional translation step for posts from these languages as a fix. The \texttt{data-errors} shadow pipeline determines how robust the user pipeline is against data quality issues such as typos by evaluating synthetic errors, and integrates an additional spell-checking step to improve the robustness. \texttt{Label-errors} uses Shapley values~\cite{jia2019efficient,karlavs2023canonpipe} to identify potentially mislabeled training samples from weak labeling, subsequently flips the label of the most likely mislabeled samples and re-evaluates the pipeline. All shadow pipelines also compute provenance-based explanations.

\headerl{Optimisations} For \texttt{slices}, SliceLine~\cite{chung2019slice,sagadeeva2021sliceline} is applied directly on the original pipeline's intermediates (parts of the unfeaturised test data and the corresponding labels) to identify the most problematic slice. In \texttt{data-errors}, we randomly corrupt 10\% of the data with typos, and recompute embeddings and run inference only for corrupted rows to measure the impact. For potential fixes, we analogously restrict expensive translation and spelling correction steps, embedding computation, and inference to relevant test subsets. \texttt{Label-errors} also changes the train data. For \texttt{rag}, we ensure that the expensive inference is re-run only for test set predictions based on train samples with flipped labels. The \texttt{train} variant of the user pipeline, unfortunately, requires neural network retraining; for that, we evaluate a special optimisation referred to as \texttt{opt-proxy}, which substitutes the neural network with a cheaper model and uses its performance difference as a proxy for the label update's impact. For subsequent shadow pipeline updates after the user's pipeline change, we re-use intermediates from the original shadow pipeline run wherever possible. In particular, we partially re-use translation, typo corruption, spell correction, and inference results.

\headerl{Data and APIs} We use a synthetic dataset with 1,000 rows, inspired by real depression detection data~\cite{nguyen2022depression}, since the original dataset cannot be shared due to its sensitive nature. For reproducibility, we fix random seeds and replay cached answers from external APIs for the LLM and translation step with artificial wait times. We repeat each experiment seven times, preceded by one warm-up run, and report the median runtimes.

\begin{figure}[t!]
    \centering
    \includegraphics[width=0.95\columnwidth]{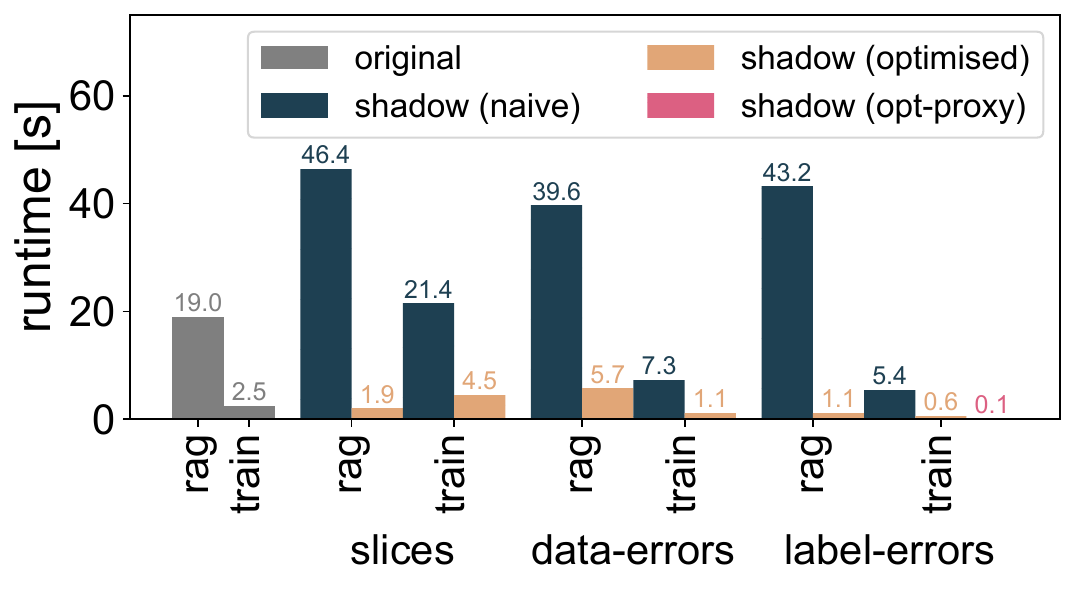} 
    \caption{Benefits of our proposed optimisations for computing shadow pipelines. Our shadow pipeline optimisations decrease the runtime by up to a factor of 38.}
    \label{fig:exp1}            
\end{figure}

\header{Computing shadow pipelines} We measure the execution time for computing the shadow pipelines after first executing the original pipeline. In each case, we compare the runtime of the \texttt{naive} baseline (repeated execution of manually rewritten versions of the original pipeline) to the runtime of the \texttt{optimised} execution, where we implement hardcoded shadow pipelines with our proposed optimisations.

\headerl{Results and discussion} Figure~\ref{fig:exp1} illustrates that the optimised shadow pipelines decrease the runtime by up to a factor of 38 (55 with the proxy model). The optimised variant consistently runs at least 4.7 times faster in all scenarios. We find that the runtime reduction primarily depends on three key factors: inference time, costly operations within shadow pipelines (e.g., spell correction, translation), and the need for retraining (not applicable for \texttt{rag}). Substituting neural network retraining with a proxy model results in a noticeable runtime improvement of 55, whereas the variant without a proxy model only achieves a speedup of nine. Notably, for the scenarios with the highest runtime (\texttt{slices} on \texttt{train} and \texttt{data-errors} on \texttt{rag}), the initial shadow pipeline execution accounts for the majority of the runtime due to expensive operators, where we can reuse intermediate results for shadow pipeline maintenance. It is noteworthy that code inspections in modern IDEs like IntelliJ can also incur several seconds of latency, which seems generally acceptable for users, with the option to disable specific inspections.

\header{Maintaining shadow pipelines} In the second experiment, we simulate that a user makes a small change to their pipeline code (the update of a regex), and then measure how long it takes to update the results of both the original pipeline and its shadow pipelines. 

\headerl{Results and discussion} The incremental shadow pipeline maintenance is up to 626 times faster than the full naive re-execution, with original pipeline runtimes decreasing up to fivefold. The additional re-use of intermediates from previous shadow pipeline executions leads to bigger runtime decreases than their initial optimised execution in the previous experiment (where we only saw decreases by up to a factor of 55). Additionally, shadow pipeline maintenance takes less than one second in all but one scenarios. The label flipping for the most likely label errors in \texttt{label-errors} on \texttt{rag} affects train rows retrieved for many test predictions, which leads to high inference costs, which we cannot avoid. Notably, we now observe particularly low latencies for the scenarios where we observed the highest runtimes in the previous experiment (\texttt{slices} on \texttt{train} and \texttt{data-errors} on \texttt{rag}), as we can reuse expensive intermediates from the initial execution of the respective shadow pipelines.

\begin{figure}[t!]
    \centering
    \includegraphics[width=0.95\columnwidth]{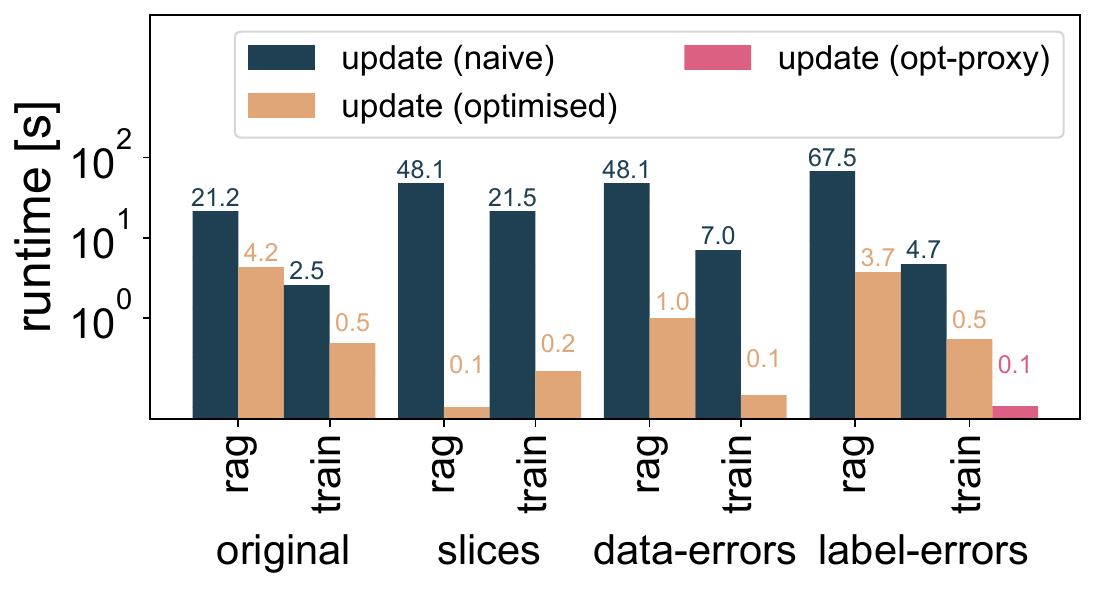} 
    \caption{Benefits of our proposed optimisations for incrementally updating the pipelines. The runtime (shown on a log scale) is less than one second in all but one scenarios.}
    \label{fig:exp2}            
\end{figure}
\section{Next Steps \& Open Questions} We plan to explore how to create and efficiently maintain shadow pipelines for user-defined pipelines, by automatically applying our outlined optimisations. In particular, we aim to develop a runtime for incremental view maintenance and shadow pipeline parallelisation (e.g., by building on DuckDB). Additionally, we plan to conduct a user study to evaluate the usefulness of our envisioned system and its provenance-based explanations for data scientists. Several questions remain open, including whether to directly build certain optimisations into the shadow pipelines or build a custom optimiser, the extent to which we will need custom incremental operators, the reliance on sampling, and approximations such as proxy models. Furthermore, we aim to explore the integration of LLM-based code suggestions \cite{githubCopilot} for detected data-related problems into our system.

\balance

\bibliographystyle{ACM-Reference-Format}
\bibliography{main}


\begin{thebibliography}{28}


\ifx \showCODEN    \undefined \def \showCODEN     #1{\unskip}     \fi
\ifx \showDOI      \undefined \def \showDOI       #1{#1}\fi
\ifx \showISBNx    \undefined \def \showISBNx     #1{\unskip}     \fi
\ifx \showISBNxiii \undefined \def \showISBNxiii  #1{\unskip}     \fi
\ifx \showISSN     \undefined \def \showISSN      #1{\unskip}     \fi
\ifx \showLCCN     \undefined \def \showLCCN      #1{\unskip}     \fi
\ifx \shownote     \undefined \def \shownote      #1{#1}          \fi
\ifx \showarticletitle \undefined \def \showarticletitle #1{#1}   \fi
\ifx \showURL      \undefined \def \showURL       {\relax}        \fi
\providecommand\bibfield[2]{#2}
\providecommand\bibinfo[2]{#2}
\providecommand\natexlab[1]{#1}
\providecommand\showeprint[2][]{arXiv:#2}

\bibitem[Bird et~al\mbox{.}(2020)]%
        {bird2020fairlearn}
\bibfield{author}{\bibinfo{person}{Sarah Bird}, \bibinfo{person}{Miro Dud{\'\i}k}, \bibinfo{person}{Richard Edgar}, \bibinfo{person}{Brandon Horn}, \bibinfo{person}{Roman Lutz}, \bibinfo{person}{Vanessa Milan}, \bibinfo{person}{Mehrnoosh Sameki}, \bibinfo{person}{Hanna Wallach}, {and} \bibinfo{person}{Kathleen Walker}.} \bibinfo{year}{2020}\natexlab{}.
\newblock \showarticletitle{Fairlearn: A toolkit for assessing and improving fairness in AI}.
\newblock \bibinfo{journal}{\emph{Microsoft, Tech. Rep. MSR-TR-2020-32}} (\bibinfo{year}{2020}).
\newblock


\bibitem[Breck et~al\mbox{.}(2019)]%
        {breck2019data}
\bibfield{author}{\bibinfo{person}{Eric Breck}, \bibinfo{person}{Neoklis Polyzotis}, \bibinfo{person}{Sudip Roy}, \bibinfo{person}{Steven Whang}, {and} \bibinfo{person}{Martin Zinkevich}.} \bibinfo{year}{2019}\natexlab{}.
\newblock \showarticletitle{Data Validation for Machine Learning}.
\newblock \bibinfo{journal}{\emph{MLSys}} (\bibinfo{year}{2019}).
\newblock


\bibitem[Chothia et~al\mbox{.}(2016)]%
        {chothia2016explaining}
\bibfield{author}{\bibinfo{person}{Zaheer Chothia}, \bibinfo{person}{John Liagouris}, \bibinfo{person}{Frank McSherry}, {and} \bibinfo{person}{Timothy Roscoe}.} \bibinfo{year}{2016}\natexlab{}.
\newblock \bibinfo{booktitle}{\emph{Explaining outputs in modern data analytics}}.
\newblock \bibinfo{type}{{T}echnical {R}eport}. \bibinfo{institution}{ETH Zurich}.
\newblock


\bibitem[Chung et~al\mbox{.}(2019)]%
        {chung2019slice}
\bibfield{author}{\bibinfo{person}{Yeounoh Chung}, \bibinfo{person}{Tim Kraska}, \bibinfo{person}{Neoklis Polyzotis}, \bibinfo{person}{Ki~Hyun Tae}, {and} \bibinfo{person}{Steven~Euijong Whang}.} \bibinfo{year}{2019}\natexlab{}.
\newblock \showarticletitle{Slice finder: Automated data slicing for model validation}. In \bibinfo{booktitle}{\emph{2019 IEEE 35th International Conference on Data Engineering (ICDE)}}. IEEE, \bibinfo{pages}{1550--1553}.
\newblock


\bibitem[GitHub(2021)]%
        {githubCopilot}
\bibfield{author}{\bibinfo{person}{GitHub}.} \bibinfo{year}{2021}\natexlab{}.
\newblock \bibinfo{title}{{GitHub Copilot · Your AI pair programmer}}.
\newblock \bibinfo{howpublished}{\url{https://copilot.github.com/}}.
\newblock


\bibitem[Grafberger et~al\mbox{.}(2023a)]%
        {grafberger2023mlwhatif}
\bibfield{author}{\bibinfo{person}{Stefan Grafberger}, \bibinfo{person}{Paul Groth}, {and} \bibinfo{person}{Sebastian Schelter}.} \bibinfo{year}{2023}\natexlab{a}.
\newblock \showarticletitle{Automating and Optimizing Data-Centric What-If Analyses on~Native~ Machine~Learning~Pipelines}.
\newblock \bibinfo{journal}{\emph{SIGMOD}} (\bibinfo{year}{2023}).
\newblock


\bibitem[Grafberger et~al\mbox{.}(2022)]%
        {grafberger2022mlinspect}
\bibfield{author}{\bibinfo{person}{Stefan Grafberger}, \bibinfo{person}{Paul Groth}, \bibinfo{person}{Julia Stoyanovich}, {and} \bibinfo{person}{Sebastian Schelter}.} \bibinfo{year}{2022}\natexlab{}.
\newblock \showarticletitle{Data distribution debugging in machine learning pipelines}.
\newblock \bibinfo{journal}{\emph{VLDBJ}} (\bibinfo{year}{2022}).
\newblock


\bibitem[Grafberger et~al\mbox{.}(2023b)]%
        {grafberger2023demo}
\bibfield{author}{\bibinfo{person}{Stefan Grafberger}, \bibinfo{person}{Shubha Guha}, \bibinfo{person}{Paul Groth}, {and} \bibinfo{person}{Sebastian Schelter}.} \bibinfo{year}{2023}\natexlab{b}.
\newblock \showarticletitle{mlwhatif: What If You Could Stop Re-Implementing Your Machine Learning Pipeline Analyses over and over?}
\newblock \bibinfo{journal}{\emph{Proc. VLDB Endow.}} \bibinfo{volume}{16}, \bibinfo{number}{12} (\bibinfo{date}{aug} \bibinfo{year}{2023}), \bibinfo{pages}{4002–4005}.
\newblock
\showISSN{2150-8097}
\urldef\tempurl%
\url{https://doi.org/10.14778/3611540.3611606}
\showDOI{\tempurl}


\bibitem[Grafberger et~al\mbox{.}(2021)]%
        {grafberger2021demo}
\bibfield{author}{\bibinfo{person}{Stefan Grafberger}, \bibinfo{person}{Shubha Guha}, \bibinfo{person}{Julia Stoyanovich}, {and} \bibinfo{person}{Sebastian Schelter}.} \bibinfo{year}{2021}\natexlab{}.
\newblock \showarticletitle{MLINSPECT: A Data Distribution Debugger for Machine Learning Pipelines}.
\newblock \bibinfo{journal}{\emph{SIGMOD}} (\bibinfo{year}{2021}).
\newblock


\bibitem[Grammarly({[n.\,d.]})]%
        {grammarlyDemo}
\bibfield{author}{\bibinfo{person}{Grammarly}.} \bibinfo{year}{[n.\,d.]}\natexlab{}.
\newblock \bibinfo{title}{{Demo}}.
\newblock \bibinfo{howpublished}{\url{https://demo.grammarly.com/}}.
\newblock


\bibitem[Holstein et~al\mbox{.}(2019)]%
        {holstein2019improving}
\bibfield{author}{\bibinfo{person}{Kenneth Holstein}, \bibinfo{person}{Jennifer Wortman~Vaughan}, \bibinfo{person}{Hal Daum{\'e}~III}, \bibinfo{person}{Miro Dudik}, {and} \bibinfo{person}{Hanna Wallach}.} \bibinfo{year}{2019}\natexlab{}.
\newblock \showarticletitle{Improving fairness in machine learning systems: What do industry practitioners need?}. In \bibinfo{booktitle}{\emph{Proceedings of the 2019 CHI conference on human factors in computing systems}}. \bibinfo{pages}{1--16}.
\newblock


\bibitem[Jetbrains({[n.\,d.]})]%
        {codeInspectionsIntelliJ}
\bibfield{author}{\bibinfo{person}{Jetbrains}.} \bibinfo{year}{[n.\,d.]}\natexlab{}.
\newblock \bibinfo{title}{{Code inspections}}.
\newblock \bibinfo{howpublished}{\url{https://www.jetbrains.com/help/idea/code-inspection.html\#access-inspections-and-settings}}.
\newblock


\bibitem[Jia et~al\mbox{.}(2019)]%
        {jia2019efficient}
\bibfield{author}{\bibinfo{person}{Ruoxi Jia}, \bibinfo{person}{David Dao}, \bibinfo{person}{Boxin Wang}, \bibinfo{person}{Frances~Ann Hubis}, \bibinfo{person}{Nezihe~Merve Gurel}, \bibinfo{person}{Bo Li}, \bibinfo{person}{Ce Zhang}, \bibinfo{person}{Costas~J Spanos}, {and} \bibinfo{person}{Dawn Song}.} \bibinfo{year}{2019}\natexlab{}.
\newblock \showarticletitle{Efficient task-specific data valuation for nearest neighbor algorithms}.
\newblock \bibinfo{journal}{\emph{PVLDB}} (\bibinfo{year}{2019}).
\newblock


\bibitem[Karla{\v{s}} et~al\mbox{.}(2023)]%
        {karlavs2023canonpipe}
\bibfield{author}{\bibinfo{person}{Bojan Karla{\v{s}}}, \bibinfo{person}{David Dao}, \bibinfo{person}{Matteo Interlandi}, \bibinfo{person}{Sebastian Schelter}, \bibinfo{person}{Wentao Wu}, {and} \bibinfo{person}{Ce Zhang}.} \bibinfo{year}{2023}\natexlab{}.
\newblock \showarticletitle{Data Debugging with Shapley Importance over Machine Learning Pipelines}. In \bibinfo{booktitle}{\emph{The Twelfth International Conference on Learning Representations}}.
\newblock


\bibitem[Li et~al\mbox{.}(2019)]%
        {li2019cleanml}
\bibfield{author}{\bibinfo{person}{Peng Li}, \bibinfo{person}{Xi Rao}, \bibinfo{person}{Jennifer Blase}, \bibinfo{person}{Yue Zhang}, \bibinfo{person}{Xu Chu}, {and} \bibinfo{person}{Ce Zhang}.} \bibinfo{year}{2019}\natexlab{}.
\newblock \showarticletitle{Cleanml: A benchmark for joint data cleaning and machine learning}.
\newblock \bibinfo{journal}{\emph{ICDE}} (\bibinfo{year}{2019}).
\newblock


\bibitem[McSherry et~al\mbox{.}(2013)]%
        {mcsherry2013differential}
\bibfield{author}{\bibinfo{person}{Frank McSherry}, \bibinfo{person}{Derek~Gordon Murray}, \bibinfo{person}{Rebecca Isaacs}, {and} \bibinfo{person}{Michael Isard}.} \bibinfo{year}{2013}\natexlab{}.
\newblock \showarticletitle{Differential Dataflow.}. In \bibinfo{booktitle}{\emph{CIDR}}.
\newblock


\bibitem[Nguyen et~al\mbox{.}(2022)]%
        {nguyen2022depression}
\bibfield{author}{\bibinfo{person}{Thong Nguyen}, \bibinfo{person}{Andrew Yates}, \bibinfo{person}{Ayah Zirikly}, \bibinfo{person}{Bart Desmet}, {and} \bibinfo{person}{Arman Cohan}.} \bibinfo{year}{2022}\natexlab{}.
\newblock \showarticletitle{Improving the Generalizability of Depression Detection by Leveraging Clinical Questionnaires}. In \bibinfo{booktitle}{\emph{Proceedings of the 60th Annual Meeting of the Association for Computational Linguistics (Volume 1: Long Papers)}}, \bibfield{editor}{\bibinfo{person}{Smaranda Muresan}, \bibinfo{person}{Preslav Nakov}, {and} \bibinfo{person}{Aline Villavicencio}} (Eds.). \bibinfo{publisher}{Association for Computational Linguistics}, \bibinfo{address}{Dublin, Ireland}, \bibinfo{pages}{8446--8459}.
\newblock
\urldef\tempurl%
\url{https://doi.org/10.18653/v1/2022.acl-long.578}
\showDOI{\tempurl}


\bibitem[Polyzotis et~al\mbox{.}(2018)]%
        {polyzotis2018data}
\bibfield{author}{\bibinfo{person}{Neoklis Polyzotis}, \bibinfo{person}{Sudip Roy}, \bibinfo{person}{Steven~Euijong Whang}, {and} \bibinfo{person}{Martin Zinkevich}.} \bibinfo{year}{2018}\natexlab{}.
\newblock \showarticletitle{Data lifecycle challenges in production machine learning: a survey}.
\newblock \bibinfo{journal}{\emph{SIGMOD Record}} \bibinfo{volume}{47}, \bibinfo{number}{2} (\bibinfo{year}{2018}).
\newblock


\bibitem[Pradhan et~al\mbox{.}(2022)]%
        {pradhan2022interpretable}
\bibfield{author}{\bibinfo{person}{Romila Pradhan}, \bibinfo{person}{Jiongli Zhu}, \bibinfo{person}{Boris Glavic}, {and} \bibinfo{person}{Babak Salimi}.} \bibinfo{year}{2022}\natexlab{}.
\newblock \showarticletitle{Interpretable data-based explanations for fairness debugging}. In \bibinfo{booktitle}{\emph{Proceedings of the 2022 International Conference on Management of Data}}. \bibinfo{pages}{247--261}.
\newblock


\bibitem[Sagadeeva and Boehm(2021)]%
        {sagadeeva2021sliceline}
\bibfield{author}{\bibinfo{person}{Svetlana Sagadeeva} {and} \bibinfo{person}{Matthias Boehm}.} \bibinfo{year}{2021}\natexlab{}.
\newblock \showarticletitle{Sliceline: Fast, linear-algebra-based slice finding for ml model debugging}. In \bibinfo{booktitle}{\emph{Proceedings of the 2021 International Conference on Management of Data}}. \bibinfo{pages}{2290--2299}.
\newblock


\bibitem[Schelter et~al\mbox{.}(2018a)]%
        {schelter2018challenges}
\bibfield{author}{\bibinfo{person}{Sebastian Schelter}, \bibinfo{person}{Felix Biessmann}, \bibinfo{person}{Tim Januschowski}, \bibinfo{person}{David Salinas}, \bibinfo{person}{Stephan Seufert}, {and} \bibinfo{person}{Gyuri Szarvas}.} \bibinfo{year}{2018}\natexlab{a}.
\newblock \showarticletitle{On challenges in machine learning model management}.
\newblock \bibinfo{journal}{\emph{IEEE Data Engineering Bulletin}} (\bibinfo{year}{2018}).
\newblock


\bibitem[Schelter et~al\mbox{.}(2023)]%
        {schelter2023demo}
\bibfield{author}{\bibinfo{person}{Sebastian Schelter}, \bibinfo{person}{Stefan Grafberger}, \bibinfo{person}{Shubha Guha}, \bibinfo{person}{Bojan Karla{\v{s}}}, {and} \bibinfo{person}{Ce Zhang}.} \bibinfo{year}{2023}\natexlab{}.
\newblock \showarticletitle{Proactively Screening Machine Learning Pipelines with ArgusEyes}.
\newblock \bibinfo{journal}{\emph{SIGMOD}} (\bibinfo{year}{2023}).
\newblock


\bibitem[Schelter et~al\mbox{.}(2022)]%
        {schelter2022screening}
\bibfield{author}{\bibinfo{person}{Sebastian Schelter}, \bibinfo{person}{Stefan Grafberger}, \bibinfo{person}{Shubha Guha}, \bibinfo{person}{Olivier Sprangers}, \bibinfo{person}{Bojan Karla{\v{s}}}, {and} \bibinfo{person}{Ce Zhang}.} \bibinfo{year}{2022}\natexlab{}.
\newblock \showarticletitle{Screening Native ML Pipelines with “ArgusEyes”}.
\newblock \bibinfo{journal}{\emph{CIDR}} (\bibinfo{year}{2022}).
\newblock


\bibitem[Schelter et~al\mbox{.}(2018b)]%
        {schelter2018deequ}
\bibfield{author}{\bibinfo{person}{Sebastian Schelter}, \bibinfo{person}{Dustin Lange}, \bibinfo{person}{Philipp Schmidt}, \bibinfo{person}{Meltem Celikel}, \bibinfo{person}{Felix Biessmann}, {and} \bibinfo{person}{Andreas Grafberger}.} \bibinfo{year}{2018}\natexlab{b}.
\newblock \showarticletitle{Automating Large-Scale Data Quality Verification}.
\newblock \bibinfo{journal}{\emph{Proc. VLDB Endow.}} \bibinfo{volume}{11}, \bibinfo{number}{12} (\bibinfo{date}{aug} \bibinfo{year}{2018}), \bibinfo{pages}{1781–1794}.
\newblock
\showISSN{2150-8097}
\urldef\tempurl%
\url{https://doi.org/10.14778/3229863.3229867}
\showDOI{\tempurl}


\bibitem[Schelter et~al\mbox{.}(2021)]%
        {schelter2021jenga}
\bibfield{author}{\bibinfo{person}{Sebastian Schelter}, \bibinfo{person}{Tammo Rukat}, {and} \bibinfo{person}{Felix Biessmann}.} \bibinfo{year}{2021}\natexlab{}.
\newblock \showarticletitle{JENGA - A Framework to Study the Impact of Data Errors on the Predictions of Machine Learning Models.}
\newblock \bibinfo{journal}{\emph{EDBT}} (\bibinfo{year}{2021}).
\newblock


\bibitem[Stoyanovich et~al\mbox{.}(2022)]%
        {stoyanovich2022}
\bibfield{author}{\bibinfo{person}{Julia Stoyanovich}, \bibinfo{person}{Bill Howe}, \bibinfo{person}{Serge Abiteboul}, \bibinfo{person}{H.V. Jagadish}, {and} \bibinfo{person}{Sebastian Schelter}.} \bibinfo{year}{2022}\natexlab{}.
\newblock \showarticletitle{Responsible Data Management}. In \bibinfo{booktitle}{\emph{Communications of the ACM}}.
\newblock


\bibitem[Tramer et~al\mbox{.}(2017)]%
        {tramer2017fairtest}
\bibfield{author}{\bibinfo{person}{Florian Tramer}, \bibinfo{person}{Vaggelis Atlidakis}, \bibinfo{person}{Roxana Geambasu}, \bibinfo{person}{Daniel Hsu}, \bibinfo{person}{Jean-Pierre Hubaux}, \bibinfo{person}{Mathias Humbert}, \bibinfo{person}{Ari Juels}, {and} \bibinfo{person}{Huang Lin}.} \bibinfo{year}{2017}\natexlab{}.
\newblock \showarticletitle{Fairtest: Discovering unwarranted associations in data-driven applications}. In \bibinfo{booktitle}{\emph{2017 IEEE European Symposium on Security and Privacy (EuroS\&P)}}. IEEE, \bibinfo{pages}{401--416}.
\newblock


\bibitem[Wu et~al\mbox{.}(2020)]%
        {wu2020complaint}
\bibfield{author}{\bibinfo{person}{Weiyuan Wu}, \bibinfo{person}{Lampros Flokas}, \bibinfo{person}{Eugene Wu}, {and} \bibinfo{person}{Jiannan Wang}.} \bibinfo{year}{2020}\natexlab{}.
\newblock \showarticletitle{Complaint-driven training data debugging for query 2.0}. In \bibinfo{booktitle}{\emph{Proceedings of the 2020 ACM SIGMOD International Conference on Management of Data}}. \bibinfo{pages}{1317--1334}.
\newblock


\end{thebibliography}

\end{document}